\documentclass{PoS}

\title{SU(2) meets SU(3) in lattice-Landau-gauge gluon and ghost propagators}

\ShortTitle{SU(2) meets SU(3) in lattice-Landau-gauge gluon and ghost 
propagators}

\author{A. Cucchieri\\
        Instituto de F\'{\i}sica de S\~ao Carlos, Universidade de S\~ao Paulo,
        \\ Caixa Postal 369, 13560-970 S\~ao Carlos, SP, Brazil
        E-mail: \email{attilio@ifsc.usp.br}}
\author{T. Mendes\\
        Instituto de F\'{\i}sica de S\~ao Carlos, Universidade de S\~ao Paulo,
        \\ Caixa Postal 369, 13560-970 S\~ao Carlos, SP, Brazil
        E-mail: \email{mendes@ifsc.usp.br}}
\author{\speaker{Orlando Oliveira} \\
        Department of Physics, University of Coimbra, 3004 516 Coimbra, 
        Portugal\\
        E-mail: \email{orlando@teor.fis.uc.pt}}
\author{P. J. Silva\\
        Department of Physics, University of Coimbra, 3004 516 Coimbra, 
        Portugal\\
        E-mail: \email{psilva@teor.fis.uc.pt}}

\abstract{A comparative study of the lattice Landau gauge gluon and ghost 
          propagators for SU(2) and SU(3) pure Yang-Mills theories is carried 
          out. The data were specially produced with equivalent lattice 
          parameters to allow for a careful comparison of the two cases. We 
          find very good agreement between the two theories. Our results seem 
          to confirm the predicton of Schwinger-Dyson equations that the 
          infrared exponents are independent of the gauge group SU(N).}

\FullConference{The XXV International Symposium on Lattice Field Theory\\
		 July 30 - August 4 2007\\
		 Regensburg, Germany}

\begin{document}

\section{Introduction and Motivation}

The investigation of the infrared limit of QCD is of central importance for the
comprehension of the mechanisms of quark and gluon confinement and of 
chiral-symmetry breaking. However, despite the recent progress, we still do
not have the full picture of the infrared structure of Yang-Mills theories.

In what concerns gluon confinement, in Landau gauge, the infrared behavior of 
gluon and ghost propagators is linked with the Gribov-Zwanziger 
\cite{Gribov,Zwanziger} and the Kugo-Ojima \cite{KugoOjima} confinement 
scenarios. These confinement mechanisms predict, at small momenta, an enhanced
ghost propagator and a suppression of the gluon propagator. Analytic studies 
of gluon and ghost propagators using Schwinger-Dyson equations (SDE) 
\cite{SmekalHA,LercheSmekal,ZwanzigerSDE} seem to agree with the above 
scenarios. Schwinger-Dyson equations are an infinite tower of nonlinear 
equations. Typically, the computation of a solution requires the definition of
a truncation scheme and the parametrization of vertices. The above mentioned
solutions are not the only known solutions. Indeed, in 
\cite{Aguilar03,Aguilar04} the authors found a set of solutions 
which do not comply with the above mechanisms. In what concerns the lattice 
results for the gluon and ghost propagators, in Landau gauge, one side 
they seem to support the analytical studies \cite{OOPJS,PJSOOPRD,PJSOOLat07}, 
on the other 
side they do not confirm the precise predictions obtained with SDE 
\cite{CucchLat07}. The solution of this apparent puzzle requires further
studies.

In the Schwinger-Dyson equations, when dynamic quarks are neglected, assuming
that $g^2 \sim 1 / N_c$ --- as suggested by analysis of the large $N_c$ limit
\cite{Hooft} --- the SDE predict that gluon and ghost propagators are 
independent of the number of colors (in the nonperturbative regime). In 
particular, they predict for the gluon and for the ghost propagators an 
infrared exponent that is independent of the gauge group $SU(N_c)$.
In this paper, we carry out a comparative study of lattice Landau gauge 
propagators for these two gauge groups. Our data were especially produced by 
considering equivalent lattice parameters in order to allow a careful 
comparison of the two cases. For details on the simulation see \cite{CMOS}.
For another study comparing $SU(2)$ and $SU(3)$ propagators see 
\cite{adelaide}.
In the following the effect of Gribov copies is not taken into account.

\section{Numerical Simulations}

We consider four different sets of lattice parameters, with the same lattice 
size $N^4$ and the same physical lattice spacing $a$ for the two gauge groups
(see Table \ref{Tsetup}). The first three cases are chosen to yield 
approximately the same physical lattice volume $V \approx (1.7$ fm$)^4$. This
allows a comparison of discretization effects. The fourth case corresponds to
a significantly larger physical volume, $V \approx (3.2$ fm$)^4$, in order to 
study finite-size effects. For all four cases, 50 configurations were 
generated using the Wilson action.The gluon and the ghost propagators
\begin{eqnarray}
   D^{ab}_{\mu\nu} (k^2) ~ = &&
       \delta^{ab} \, \left( \delta_{\mu\nu} - \frac{k_\mu k_\nu}{k^2} \right)
                   \, D( k^2 )  \, , \\
   G^{ab} (k^2) ~ = && - \delta^{ab} \, G( k^2 )
\end{eqnarray}
were computed for four different types of momenta: $(k,0,0,0)$, $(k,k,0,0)$,
$(k,k,k,0)$ and $(k,k,k,k)$. In the computation of $D(k^2)$ and $G(k^2)$, an
average over equivalent momenta and color components was always performed.

\begin{table}
\begin{center}
\begin{tabular}{lllll}
 $N^4$    & $a$ (fm)  & $Na$ (fm)  & $\beta_{SU(2)}$  & $\beta_{SU(3)}$ \\
\hline
  $16^4$  &  $0.102$  &  $1.632$  &  $2.4469$  &  $6.0$  \\
  $24^4$  &  $0.073$  &  $1.752$  &  $2.5501$  &  $6.2$  \\
  $32^4$  &  $0.054$  &  $1.728$  &  $2.6408$  &  $6.4$  \\
  $32^4$  &  $0.102$  &  $3.264$  &  $2.4469$  &  $6.0$  \\
\hline
\end{tabular}
\caption{\label{Tsetup} Lattice setup. The lattice spacing was computed
from the string tension, assuming $\sqrt{\sigma} = 440$ MeV.}
\end{center}
\end{table}

In order to compare the propagators from the different simulations, the gluon 
and ghost propagators were renormalized accordingly to
\begin{equation}
   \left. D(q^2) \right|_{q^2 = \mu^2} ~ = ~ \frac{1}{\mu^2}, \hspace{1cm} 
   \left. G(q^2) \right|_{q^2 = \mu^2} ~ = ~ \frac{1}{\mu^2},
\end{equation}
using $\mu = 3$ GeV as a renormalization point. The lattice data were
interpolated (using splines) to allow the use of such a renormalization point
in all the simulations.

\section{The Propagators}

\begin{figure}
\begin{center}
\includegraphics[scale=0.45]{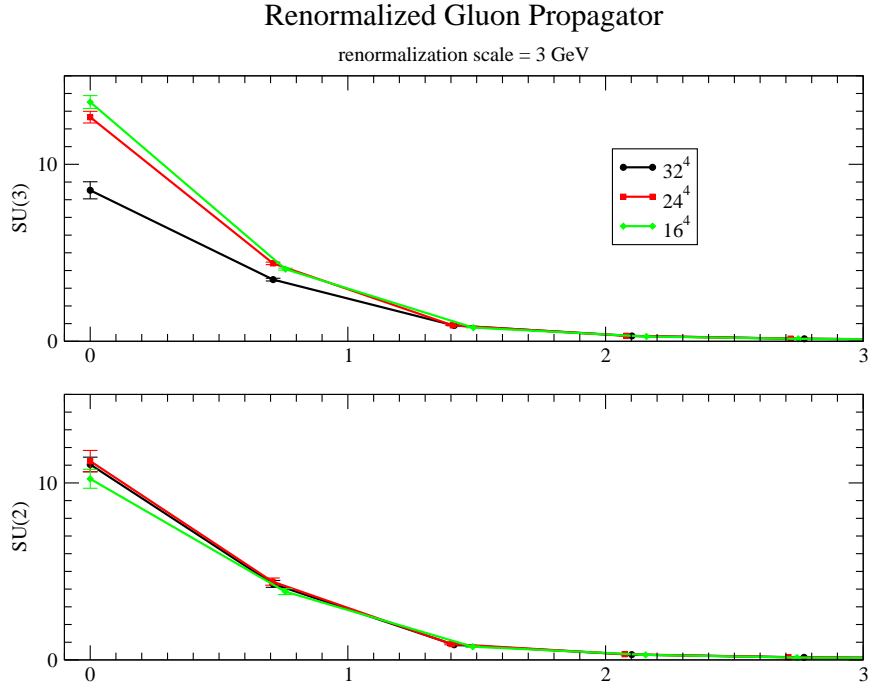}
\caption{Gluon propagator as function of momenta given in GeV
for lattices with volume $V \approx (1.7$ fm$)^4$.
\label{glue1}}
\end{center}
\end{figure}

\begin{figure}
\begin{center}
\includegraphics[scale=0.45]{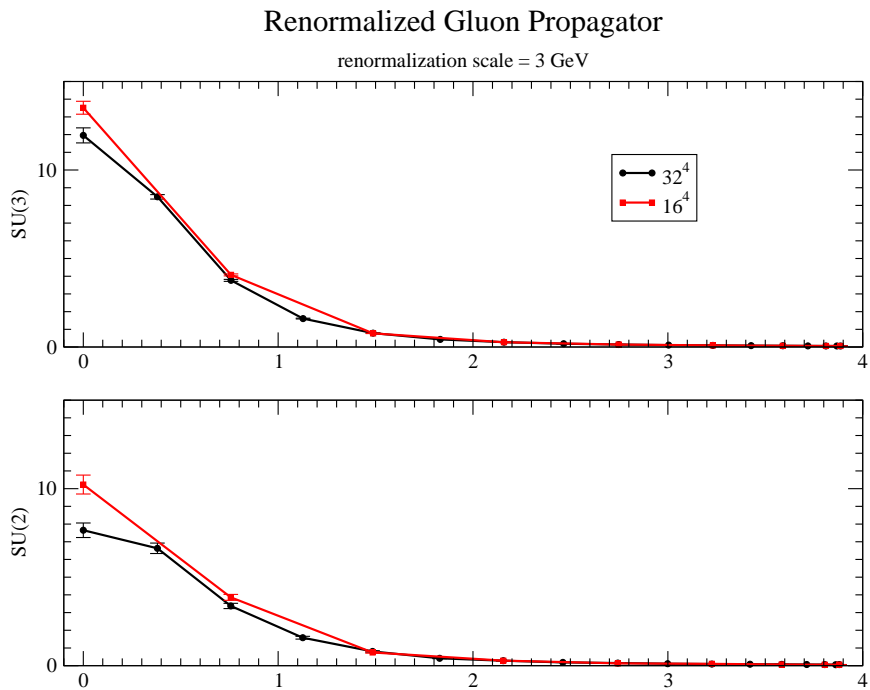}
\caption{Gluon propagator as function of momenta given in GeV for 
$\beta_{SU(3)} = 6.0$ and 
$\beta_{SU(2)} = 2.4469$. \label{glue2}}
\end{center}
\end{figure}

\begin{figure}
\begin{center}
\includegraphics[scale=0.45]{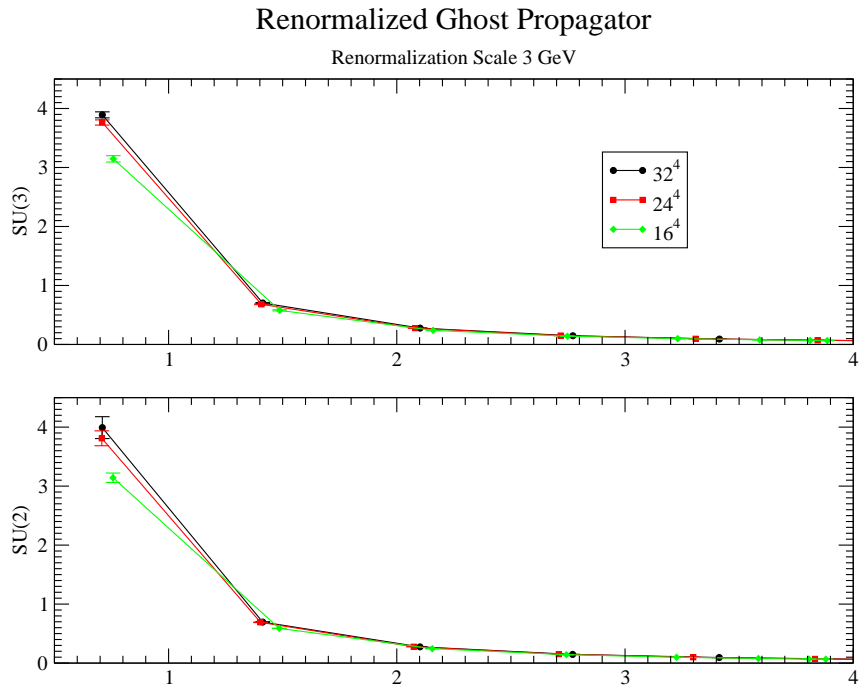}
\caption{Ghost propagator as function of momenta given in GeV 
for lattices with volume $V \approx (1.7$ fm$)^4$.
\label{ghost1}}
\end{center}
\end{figure}

\begin{figure}
\begin{center}
\includegraphics[scale=0.45]{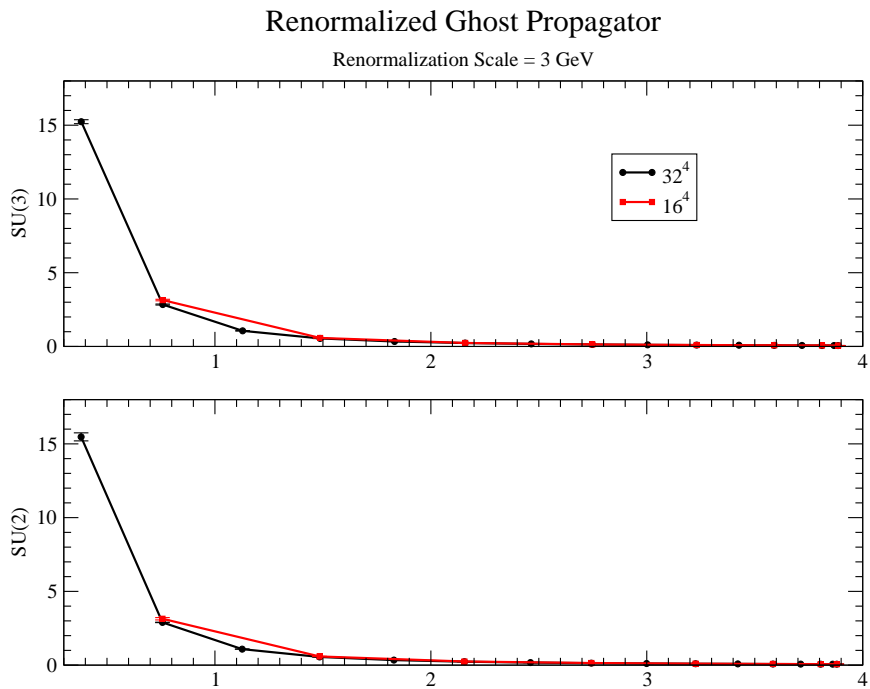}
\caption{Ghost propagator as function of momenta given in GeV 
for $\beta_{SU(3)} = 6.0$ and 
$\beta_{SU(2)} = 2.4469$. \label{ghost2}}
\end{center}
\end{figure}

The gluon propagator for $V \approx (1.7$ fm$)^4$ is reported in figure
\ref{glue1}. In figure \ref{glue2}, the data for different volumes, same
$\beta$ value is displayed. The corresponding figures for the ghost
propagator are fig. \ref{ghost1} and fig. \ref{ghost2}, respectively.


\begin{figure}
\begin{center}
\includegraphics[scale=0.45]{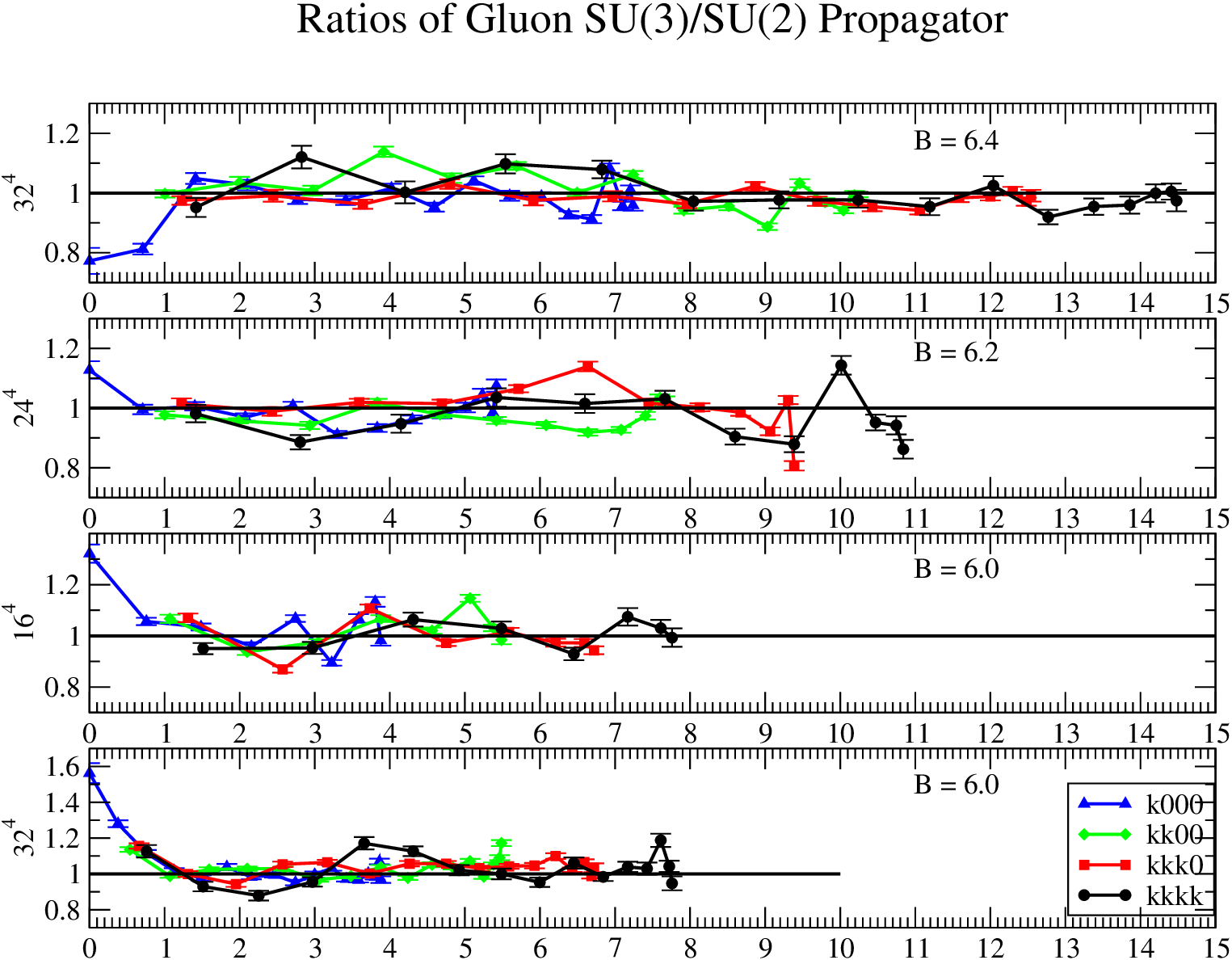}
\caption{SU(3)/SU(2) gluon propagator as function of momenta given in GeV. 
\label{ratioglue}}
\end{center}
\end{figure}

For the gluon propagator, figure \ref{glue1} show some discretization effects
which are stronger for the SU(3) data. On the other hand, figure \ref{glue2}
shows finite volume effects, specially for SU(2). In order to try to
understand such effects, in figure \ref{ratioglue} one plots the ratios of 
SU(3) over SU(2) propagators for all the simulations. Note that the plots 
include ratios of $D(0)$, i.e. the most left point should be taken with care.
In what concerns the gluon propagator, given the relatively small statistics 
and given that there is no clear systematics in data, one can not conclude on 
the nature of observed small differences. Anyway, the SU(3) and SU(2) 
propagators are, at least, qualitatively similar. Given the small differences 
one can also claim quantitative agreement between the two propagators.


\begin{figure}
\begin{center}
\includegraphics[scale=0.45]{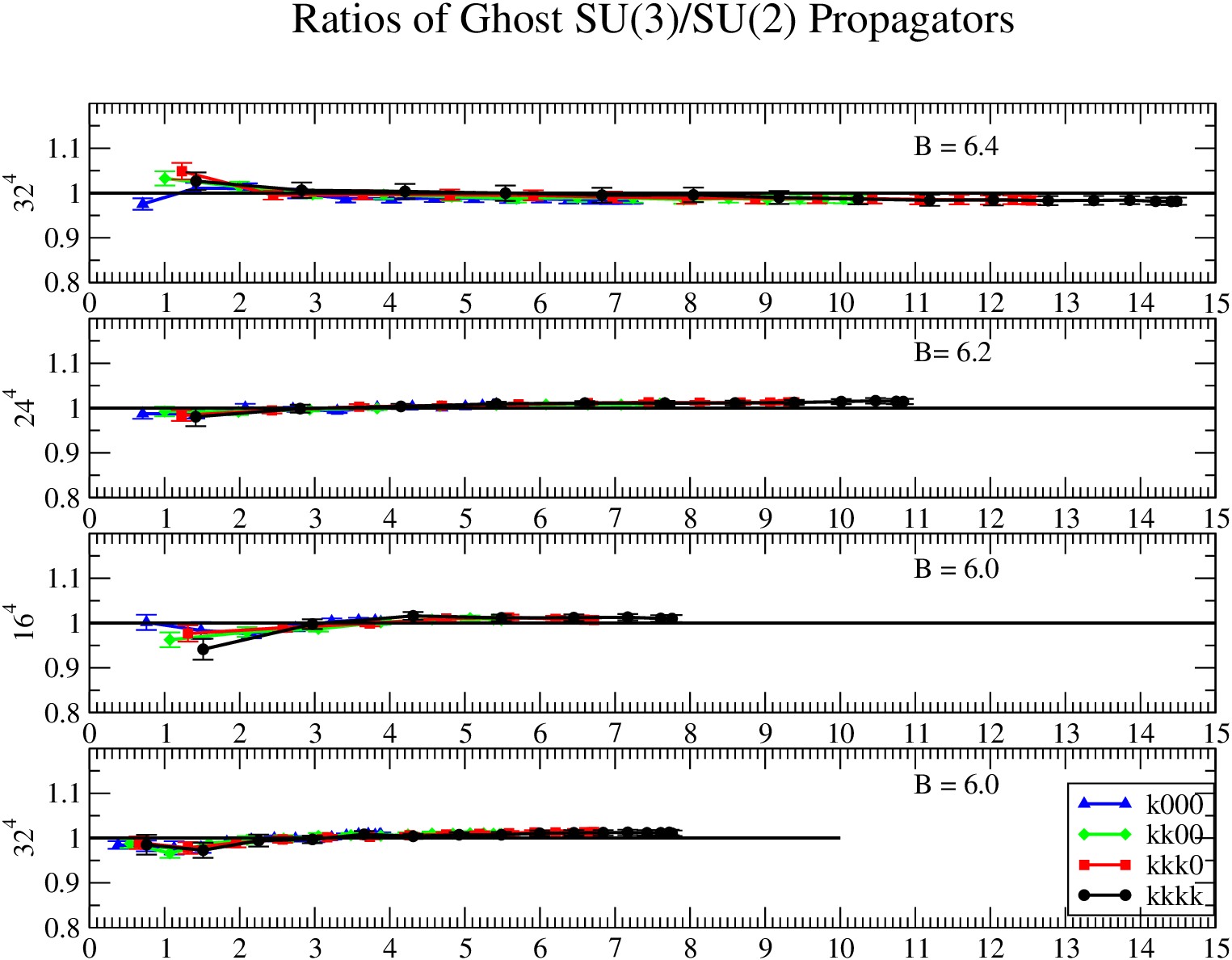}
\caption{SU(3)/SU(2) ghost propagator as function of momenta given in GeV.
 \label{ratioghost}}
\end{center}
\end{figure}

In what concerns the ghost propagator, the data seems more stable than
the gluon points. Indeed, comparing figures \ref{glue1}-\ref{ghost2} and the
ratios of propagators in fig. \ref{ratioglue}, fig \ref{ratioghost} the
ghost data fluctuates less. Moreover, for the full range of momenta the ratios
of ghost propagators are compatible with one at the level of two standard 
deviations. Therefore, for the ghost propagator one can conclude in favour
of quantitative and qualitative agreement between SU(2) and SU(3).


\section{Results and Conclusions}

In summary, considering a careful choice of the lattice parameters, 
we were able to carry out an unambiguous comparison of the lattice Landau gluon
and ghost propagators for $SU(2)$ and $SU(3)$ gauge theories. The data
show that the two cases have very similar finite-size and discretization
effects. Moreover, we find very good agreement between the two Yang-Mills 
theories (for our values of momenta larger than 1 GeV),
for all lattice parameters and for all types of momenta. Below 1 GeV, 
the results for the two gauge groups show some differences, especially 
for the gluon propagator. 
However, given the lattice volumes considered, further studies
are required before drawing conclusions about the comparison between $SU(2)$
and $SU(3)$ propagators in the deep-IR region.
In this sense, we claim that our results support the prediction from
the Schwinger-Dyson equations
that the propagators are the same for all $SU(N_c)$ groups in the
nonperturbative region.

\acknowledgments

P.J.S.acknowledges F.C.T. financial support via grant SFRH/BD/10740/2002.
This work was supported in part by F.C.T. under contracts POCI/FP/63436/2005
and POCI/FP/63923/2005.


\end{document}